\documentclass[prb,preprint,onecolumn,superscriptaddress,showpacs]{revtex4}

\usepackage{latexsym}
\usepackage{graphicx}
\usepackage{braket}

\begin{document}

\author{Z. Kim}
\author{V. Zaretskey}
\affiliation{Laboratory for Physical Sciences, College Park,
Maryland, 20740} \affiliation{Department of Physics, University of
Maryland, College Park, Maryland, 20742}
\author{Y. Yoon}
\affiliation{Department of Physics, University of Maryland,
College Park, Maryland, 20742}
\author{J. F. Schneiderman}
\author{M. D. Shaw}
\affiliation{Department of Physics, University of Southern
California, Los Angeles, California 90089-0484}
\author{P. M. Echternach}
\affiliation{Jet Propulsion Laboratory, California Institute of
Technology, Pasadena, California, 91109}
\author{F. C. Wellstood}
\affiliation{Department of Physics, University of Maryland,
College Park, Maryland, 20742}
\affiliation{Joint Quantum Institute and Center for Nanophysics and Advanced Materials}
\author{B. S. Palmer}\email{bpalmer@lps.umd.edu}
\affiliation{Laboratory for Physical Sciences, College Park,
Maryland, 20740}

\title{Anomalous avoided level crossings in a Cooper-pair box spectrum}

\newcommand{\etal}{\textit{et al.}}
\newcommand{\kb}{k_{B}}

\date{\today}

\begin{abstract}
We have observed a few distinct anomalous avoided level crossings
and voltage dependent transitions in the excited state spectrum of
an Al/AlO$_{\mbox{x}}$/Al Cooper-pair box (CPB). The device was
measured at 40 mK in the 15 - 50 GHz frequency range. We find that
a given level crosses the CPB spectrum at two different gate
voltages; the frequency and splitting size of the two crossings
differ and the splitting size depends on the Josephson energy of
the CPB. We show that this behavior is not only consistent with
the CPB being coupled to discrete charged ``two-level'' quantum
systems which move atomic distances in the CPB junctions but that
the spectra provide new information about the fluctuators, which
is not available from phase qubit spectra of anomalous avoided
levels. In particular by fitting a model Hamiltonian to our data,
we extract microscopic parameters for each fluctuator, including
well asymmetry, tunneling amplitude, and the minimum hopping
distance for each fluctuator. The tunneling rates range from less
than 3.5 to 13 GHz, which represent values between 5\% and 150\%
of the well asymmetry, and the dipole moments give a minimum
hopping distance of 0.3 to 0.8~{\AA}. We have also found that
these discrete two-level systems have a pronounced effect on the
relaxation time ($T_{1}$) of the quantum states of the CPB and
hence can be a source of dissipation for superconducting quantum
bits.
\end{abstract}

\pacs{85.25.Cp and 85.35.Gv} \maketitle

Two-level fluctuators (TLF) of one kind or another have long been
recognized as an underlying cause of resistance fluctuations in
metals,~\cite{Dutta1979,Rogers1985} critical current fluctuations
in Josephson
junctions,~\cite{Constantin2007,Michael2005,Wellstood2004,Harlingen2004,Koch1983}
excess flux noise in
SQUIDs,~\cite{Koch1983,Chiorescu2003,Pashkin2003} and telegraph
noise and excess charge noise in Coulomb blockade
devices.~\cite{Geerligs1990,Zimmerli1992,Ji1994,Kenyon2000} For
charge and critical current fluctuations - the fluctuators appear
to be moving charges or rotating electric dipoles in the
insulating junction barrier or nearby dielectric
layers~\cite{Constantin2007} - but the precise microscopic
identification of the fluctuators is not settled.

There are several reasons why it has been difficult to be certain
of the precise microscopic agents causing charge or critical
current fluctuations. First, the effects from individual
fluctuators are typically very small, especially for the cryogenic
or milli-Kelvin temperatures of interest here, and this typically
makes the resulting experiments quite challenging. Second, a
variety of materials and processes have been used to build
devices, and the presence of a fluctuator may well depend on both
the materials and the fabrication technique. Third, while
measurements of telegraph noise can reveal microscopic information
about individual fluctuators, in many cases it is possible only to
distinguish the largest one or two fluctuators in a background
noise of smaller fluctuations. Compared to the number of atoms in
even the smallest device, such observable discrete fluctuators are
extremely rare and thus may not be representative of a typical
atomic scale defect in the device. Finally, much of the
experimental data obtained on fluctuators consists of relatively
smooth 1/f noise power spectra. Such noise spectra arise from a
distribution of many fluctuators~\cite{Dutta1979,Machlup1954} and
it is often not possible to resolve individual fluctuators. From
smooth spectra, it is difficult to determine uniquely such basic
microscopic parameters as the hopping distance or the absolute
number of fluctuators in a given energy range.

Research in quantum computing based on superconducting devices has
led to increased interest in understanding two-level fluctuators.
In qubit research, fluctuators can be a serious problem because
they can lead to decoherence, dissipation, inhomogenous broadening
and a decrease in measurement
fidelity.~\cite{SimmondsPRL2004,Martinis2005,TianPRL2007,PlourdePRB2005,Ithier2005,DeppePRB2007}
With this new interest have also come new approaches to the
problem. In particular, microwave spectroscopy of the various
superconducting qubits has revealed the presence of small
un-intended avoided crossings in the transition spectrum due to
coupling between the device and individual two-level
fluctuators.~\cite{SimmondsPRL2004,Martinis2005,PlourdePRB2005,DeppePRB2007,Ithier2005,SchreierPRB2008}
Avoided level crossings were initially seen in the phase
qubit~\cite{SimmondsPRL2004} and subsequently in the flux
qubit~\cite{PlourdePRB2005,DeppePRB2007},
quantronium~\cite{Ithier2005}, and the
transmon.~\cite{SchreierPRB2008} While it was initially thought
that the anomalous avoided level crossings were due to critical
current fluctuations,~\cite{SimmondsPRL2004} a later comparison of
a hopping distance (extracted from an analysis of an ensemble of
avoided level crossings) to atomic distances suggested that charge
fluctuators were responsible.~\cite{Martinis2005}

In this article, we report observations of anomalous avoided level
crossings in the transition spectrum of a Cooper-pair box (CPB), a
superconducting quantum device that is sensitive to charge. We
find that each avoided level crossing is due to a charged two
level system in which the charge moves atomic distances in the
tunnel barrier of the device's Josephson junction. The spectra
contain a striking feature not seen in phase qubits - the two
level system's levels are voltage dependent - and by modeling the
full spectrum, we can resolve some of the key microscopic
parameters of each fluctuator. We also find a strong correlation
between the locations of the avoided level crossings and decreases
in the lifetime ($T_{1}$) of the excited state of the qubit.

For our measurement, we use a CPB consisting of a small
superconducting island connected to superconducting leads by two
ultra-small (nominal area 120 nm by 120 nm)  Josephson junctions
(see Fig.~\ref{fig:ExpSetup}). By applying voltage $V_{g}$ to a
gate electrode, which is capacitively coupled to the CPB island
with capacitance $C_{g}$, we can control the number $n$ of excess
Cooper pairs on the island. Neglecting the quasiparticle
states,~\cite{PalmerPRB2007,AumentadoPRL2004} the Hamiltonian
describing the CPB is given
by~\cite{ButtikerPRB1987,BouchiatPS1998}
\begin{equation}
\hat{H}_{CPB} = E_{c} \sum_{n} (2n-n_{g})^{2}\ket{n}\bra{n} -
\frac{E_{J}}{2}\sum_{n}(\ket{n+1}\bra{n}+\ket{n}\bra{n+1})
\label{eq:HamCPB}
\end{equation}
where $n_{g}= -C_{g}V_{g}/e$ is the reduced gate voltage,
$e=-1.609\times10^{-19}$~C is negative, $E_{c} =
e^{2}/2C_{\Sigma}$ is the charging energy, $C_{\Sigma}$ is the
total capacitance of the island, and $E_{J}$ is the Josephson
energy. At the CPB charge degeneracy point, which occurs when
$n_{g}$ is an odd integer, the energy difference between the first
excited state and ground state is a minimum and approximately
$E_{J}$. By applying a magnetic flux ($\phi$) through the
superconducting loop, $E_{J}$ can be effectively changed by the
quantity $E_{J} = E_{J}^{Max} \cos(\pi \phi/\phi_{\circ})$ where
$E_{J}^{Max} = \hbar I_{c}/2e$, $I_{c}$ is the critical current of
the two junctions, and $\phi_{\circ} = hc/2e$ is the magnetic flux
quantum. Similarly, the first and second excited state have a
minimum splitting of approximately $E_{J}^{2}/16E_{c}$ and this
minimum occurs at even values of $n_{g}$.

To measure the excess charge on the CPB, we used a superconducting
Coulomb-blockade electrometer that is capacitively coupled to the
CPB [see
Fig.~\ref{fig:ExpSetup}(a)].~\cite{BladhNewJournalPhysics2005} The
electrometer and the CPB were formed together on a quartz
substrate in the same process; we used electron beam lithography
and standard double-angle evaporation of Al with an oxidation step
to form an AlO$_{\mbox{x}}$ tunnel barrier between the two Al
layers.~\cite{Fulton1987} To improve the bandwidth of the
measurement, the electrometer was connected to an on-chip LC tank
circuit and operated in an rf-SET
mode;~\cite{SchoelkopfScience1998} the reflectance was measured at
the resonance frequency of the tank circuit (640 MHz) while the
electrometer was dc biased at the Josephson quasiparticle
resonance and with the gate of the electrometer tuned to maximize
the sensitivity to charge changes.~\cite{FultonPRL1989}

The sample was mounted in a Cu box that was bolted to the mixing
chamber of an Oxford Instruments model 100 dilution refrigerator
with a base temperature of 40 mK. Figure~\ref{fig:Spectroscopy}
shows a plot of the measured change in the rf reflectance of the
rf-SET as a function of $n_{g}$ and frequency of a microwave
perturbation applied to the gate of the CPB. For this plot, the
microwave frequency was adjusted between 24 and 50 GHz in steps of
30 MHz. When the applied frequency is resonant with a level
splitting of the CPB, the CPB state changes, causing a change in
the charge on the island, which causes a change in the reflectance
of the rf-SET. The white parabolic like band between $ 0 < n_{g}
<1$ in Fig.~\ref{fig:Spectroscopy} corresponds to a measured
change of $\approx 1e$ on the island of the CPB while the black
parabola between $ 1 < n_{g} <2$ corresponds to a measured change
of $\approx - 1e$ on the island of the CPB.

Close examination of the spectrum reveals a few small avoided
crossings which are not predicted by Eq.~\ref{eq:HamCPB};
anomalous avoided crossings imply that additional degrees of
freedom are coupled to the CPB. Figure~\ref{fig:splitting} shows a
detailed view of the spectrum near two prominent avoided crossings
due to a single fluctuator for four different values of $E_{J}$.
In Fig.~\ref{fig:splitting}(a), one crossing occurs at $f=34.3$
GHz and $n_{g}=-0.43$. A second crossing occurs at $f=33.5$ GHz
and $n_{g}=0.48$. We note in particular the presence near each
crossing of short features in the spectrum that appear to point
toward the other crossing (see arrows in Fig. 3a which indicate
small projections in the spectrum), suggesting the two crossings
are related. When $n_{g}$ was swept over multiple periods, we
found that this spectrum was periodic with period 2.
Figures~\ref{fig:splitting}(a) through~\ref{fig:splitting}(d) show
that the spectrum changes as $E_{J}/k_{B}$ is decreased from 1.0 K
to 0.1 K. In particular as $E_{J}$ is reduced, the avoided
crossing splitting decreases from a size of 150 MHz to a size that
is too small for us to determine. The other anomalous avoided
crossings we observed showed similar behavior.

The dependence of the crossings on gate voltage and Josephson
energy are consistent with the CPB being coupled to a charged
fluctuator that is moving in one of the Al/AlO$_{\mbox{x}}$/Al
ultra-small junction barriers. In particular, because the avoided
level crossings have a periodicity of 2 in reduced gate voltage,
this suggests that the charge fluctuator resides in one of the two
tunnel junctions that form the CPB; the electrostatic potential of
the CPB island is given by $V_{i} = e/C_{\Sigma}(2n-n_{g})$ which
has a periodicity of 2 in $n_{g}$ for the different states of the
CPB.

In general, the Hamiltonian for a charged fluctuator would depend
on the local environment including other charged defects in the
system. To simplify the problem, we assume that the fluctuator
takes the form of a point charge moving between two potential
wells. In this approximation the Hamiltonian for the TLF is
\begin{equation}
\hat{H}_{TLF}
=\varepsilon_{a}\ket{x_{a}}\bra{x_{a}}+\varepsilon_{b}\ket{x_{b}}
\bra{x_{b}}+T_{ab}(\ket{x_{a}}\bra{x_{b}}+\ket{x_{b}}\bra{x_{a}})
\label{eq:TLF}
\end{equation}
where $\varepsilon_{a}$ is the energy of the fluctuator at
position $x_{a}$, $\varepsilon_{b}$ is the energy of the
fluctuator at position $x_{b}$, and $T_{ab}$ is the tunneling
matrix element between the two sites. For an isolated fluctuator,
the difference in energy between the two states of the TLF is
given by $\sqrt{(\varepsilon_{b}-\varepsilon_{a})^{2} +
4T_{ab}^{2}}$.

Assuming the fluctuator is a point charge that resides in the
tunnel junction, with charge $Q_{TLF}$, the change in the induced
polarization charge on the island of the CPB when the fluctuator
tunnels from position $x_{a}$ to $x_{b}$ is given by $\Delta
Q_{pi} = Q_{TLF}(x_{b} - x_{a})\cos(\eta)/d$ where $\eta$ is the
angle of displacement relative to the electric field in the
junction and $d$ is the thickness of the tunnel
junction.~\cite{JacksonEM} Since the charge on the island is
constant when $n$ is constant, a corresponding change in the
polarization charge ($-\Delta Q_{pi}$) must exist over the rest of
the island. This in turn induces a change in the induced surface
charge on the gate electrode given by $\Delta Q_{g} = \Delta
Q_{pi} C_{g}/C_{\Sigma}$. Finally, we find that the change in the
electrostatic potential of the island due to motion of the point
charge is given by~\cite{WellstoodPreprint2007}
\begin{equation}
\Delta V_{i} = \frac{Q_{TLF}}{C_{\Sigma}}\frac{(x_{b} -
x_{a})\cos(\eta)}{d}.
\end{equation}
Note that the total electrostatic potential of the island depends
on the number of excess Cooper-pairs on the island, the
displacement of the point charge and the reduced gate voltage.
Accounting for the electrostatic charging energy and the work done
by gate voltage source when the point charge moves, the following
coupling Hamiltonian is found~\cite{WellstoodPreprint2007}
\begin{equation}
\hat{H}_{CPB-TLF} = 2 E_{c}\sum_{x=x_{a},x_{b}}\sum_{n}
(2n-n_{g})\frac{Q_{TLF}}{e}\frac{x\cos(\eta)}{d}\ket{n}\ket{x}\bra{x}\bra{n}.
\label{eq:couplingHam}
\end{equation}

Combining Eqs.~\ref{eq:HamCPB},~\ref{eq:TLF},
and~\ref{eq:couplingHam} we find the total Hamiltonian for a CPB
coupled to a single fluctuator. From this, we can find the energy
levels and the transition frequencies from the ground state to the
excited states of the system and fit the measured spectrum. We
note that in this model, an avoided crossing occurs when the first
excited state of the TLF is resonant with the first excited state
of the CPB. The change in the induced polarization charge on the
CPB island due to the fluctuator being excited ultimately gives
rise to the avoided level crossings at two different frequencies
and two different reduced gate voltages (i.e. it breaks the
symmetry of the CPB). The splitting size in our model depends on
$E_{J}$; coupling the two excited states together is a second
order process that involves both the tunneling of a Cooper-pair
and the tunneling of the TLF. The dashed red curve in
Fig.~\ref{fig:splitting} shows the predicted spectrum for the
splitting we found near 34 GHz. Here we used fit parameters
$(\varepsilon_{b}-\varepsilon_{a})/\kb$ = 1.427 K, $T_{ab}/\kb$ =
0.39 K, and $Q_{TLF}\delta x\cos(\eta)/ed = -0.074$ where $\delta
x = (x_{b}-x_{a})$ is the maximum displacement of the fluctuator.
We note that extracting these three parameters requires us to
measure the splitting size and frequency of the avoided crossings
at two different gate voltages and that these fit parameters can
be varied by approximately 10\% to maintain a good fit.

We found another prominent avoided crossing near 43 GHz and
smaller avoided crossings near 20 and 39 GHz.
Table~\ref{tab:Expparameters} summarizes the best fit results for
all of the observed avoided crossings. The avoided crossing near
39 GHz had a splitting size too small to resolve, which places an
upper bound on $T_{ab}$ of the TLF. Assuming a TLF charge of
$Q_{TLF} = e$ and a tunnel barrier thickness of $d = 1$ nm, we
extract minimum hopping distances for the fluctuators that range
from 0.32 {\AA} to 0.83 {\AA}.

The model also predicts state transitions of the TLF at gate
voltages away from the observed avoided crossings. Although these
were not visible in Fig.~\ref{fig:splitting}, we found that weak
transitions could be observed far from the avoided crossing when
the microwave excitation power was increased by approximately a
factor of ten. Figure~\ref{fig:excesscharge}(a) shows the measured
excess charge spectrum between an applied frequency of 33 and 34.6
GHz; note the very faint transition due to the TLF  between
$n_{g}$~=~-1.25 and $n_{g}$~=~-0.75. In this region the measured
excess charge on the CPB is approximately -0.03 e, about an order
of magnitude smaller than predicted from our simple theory.
Figure~\ref{fig:excesscharge}(b) overlays the predicted spectrum
using the parameters found in Fig.~\ref{fig:splitting}.

Similarly, Fig.~\ref{fig:excesscharge}(c) shows the measured
excess charge spectrum between 41.5 to 44.5 GHz; a frequency range
where another prominent avoided crossing was observed.  Note that
in Fig.~\ref{fig:excesscharge}(c) we observe weak voltage
dependent transitions, due to the TLF, near $n_{g}= 0.8$ and $f =
44.25$ GHz; as well as between $n_{g} = -0.8$ and -0.4. Again the
predicted spectrum in Fig.~\ref{fig:excesscharge}(d) follows the
measured spectrum accurately. In this spectrum, we observe
additional features like non-equilibrium quasiparticles which
creates the apparent periodicity in $n_{g}$ of
1,~\cite{PalmerPRB2007, AumentadoPRL2004}  multiple CPB spectra
(i.e. spectra parallel to the CPB spectrum) which is interpreted
to be due to low frequency charge fluctuators, and horizontal
bands which are due to a change in the gain of the
Coulomb-blockade electrometer.

We have also measured the lifetime ($T_{1}$) of the excited state
of the CPB as a function of the transition frequency. For these
measurements, the CPB was excited at its zero to one transition
frequency, and the charge on the island was continuously monitored
as a function of time using the rf-SET after the excitation source
was turned off. The measurements were done at a small Josephson
energy to decouple the CPB from charge perturbations in the system
and hence increase the lifetime of the qubit to a maximum
value.~\cite{AstafievPRL2004,SchoelkopfQuantumNoise2003}
Figure~\ref{fig:decayrate} shows the measured decay rate
($\Gamma_{1}$) from 15 GHz to 45 GHz at $E_{J}/\kb=0.1$ K. Several
peaks in the decay rate are evident; we find that when the CPB is
in resonance with the TLF, the measured lifetime decreases from a
few microseconds to 1 $\mu$s or less. This decrease in $T_{1}$
near a resonance was a useful tool for finding some of the avoided
crossings. This behavior suggests that the lifetime of these TLFs
is smaller than a few microseconds and that the interaction of the
CPB with a charged TLF is a source of dissipation for the
CPB.~\cite{CooperPRL2004}

Finally, to investigate the stability of the TLFs, the device was
warmed up to room temperature. After ``annealing'' at room
temperature for 14 days, the device was cooled again to 40 mK. We
found an avoided crossing associated with one level in the 20 - 50
GHz range and one peak in the decay rate at the same frequency.
The new crossings occurred around $f = 23$ GHz and fitting to the
charge model gave: $(\varepsilon_{b}-\varepsilon_{a})/\kb$ = 0.34
K, $T_{ab}/\kb$ = 0.52 K, and $Q_{TLF}\delta x \cos(\eta)/ ed =
0.078$. We also note that the fluctuators observed here were
stable after annealing at 4 K for two days or after placing the
device in the normal state by applying a 1 T magnetic field for
one hour.

In conclusion, we have observed unintended voltage-dependent
transitions and avoided level crossings in the excited state
spectrum of a CPB, consistent with a charge fluctuator that can
tunnel between two locations separated by atomic scale distances
in the tunnel junction. Finally, we note that the spectra allow us
to extract key microscopic parameters of the TLF, such as the
tunneling matrix element, and test some theories of fluctuators in
these devices.~\cite{Constantin2007,Faoro2005,ZagoskinPRL2006}

\acknowledgments{This research was supported by the National
Security Agency. The authors would like to thank E. Tiesinga, R.
Simmonds and K. Osborn for many useful discussions of microstates.
FCW would like to acknowledge support from the Joint Quantum
Institute and the State of Maryland through the Center for
Nanophysics and Advanced Materials.}

\clearpage
\begin{figure}
\caption{(a) Schematic of experimental setup. A rf
Coulomb-blockade electrometer measures the charge state and energy
spectrum of the Cooper-pair box (CPB). The CPB and electrometer
had a charging energy of $E_{c}/k_{B}$=0.575 K and
$E_{c,E}/k_{B}$=1.1 K, respectively. (b) Scanning electron
micrograph of the device.} \label{fig:ExpSetup}
\end{figure}

\begin{figure}
\caption{Measured spectrum of CPB when the applied microwave
frequency $f$ is adjusted from 24 GHz to 50 GHz with
$E_{J}/k_{B}$= 1.12 K. The white parabolic like band between $ 0 <
n_{g} <1$ corresponds to a measured change of $\approx 1e$ on the
island of the CPB while the black parabola between $ 1 < n_{g} <2$
corresponds to a measured change of $\approx - 1e$ on the island
of the CPB.} \label{fig:Spectroscopy}
\end{figure}

\begin{figure}
\caption{Measured spectrum of CPB around 34 GHz at different
Josephson energies as specified in the graphs. The red and blue
colors represent the charge on the island corresponding to the
excited and ground state of the system, respectively. The arrows
in (a) indicate a small projection of the avoided level that
appear to point toward one another. The red dashed line is the
predicted spectrum using a Hamiltonian consisting of a charged two
level fluctuator (see Table~\ref{tab:Expparameters} for the
microscopic parameters of the TLF) coupled to a CPB. }
\label{fig:splitting}
\end{figure}

\begin{figure}
\caption{Measured excess charge spectrum at relatively large
microwave drive amplitudes near two of the more prominent avoided
level crossings. (a) Measured spectrum between 33 and 34.6 GHz
with $E_{J}/k_{B}$= 0.95 K (same avoided level crossing as in
Fig.~\ref{fig:splitting}). (b) Same as (a) with the predicted
spectrum from our system Hamiltonian (red dashed curve) using the
parameters in Fig.~\ref{fig:splitting} and
Table~\ref{tab:Expparameters} and observed local peaks in the
spectrum due to the TLF as the blue points. (c) Measured spectrum
between 43 and 44.5 GHz at $E_{J}/k_{B}$= 0.1 K. (d) Same as (c)
with predicted spectrum (red dashed curve) using the parameters in
Table~\ref{tab:Expparameters} and observed local peaks in the
spectrum due to the TLF as the blue
points.}\label{fig:excesscharge}
\end{figure}

\begin{figure}
\caption{(Color online) Measured decay rate ($\Gamma_{1}$) at
$E_{J}/\kb = 0.1$ K as a function of energy level separation after
first cooldown (blue triangles) and after the device was warmed up
to room temperature and cooled back down to $T = 40$ mK (red
circles). The observed peaks in $\Gamma_{1}$ before annealing (20,
34, 39, and 43 GHz) and after annealing (23 GHz) correlate with
the location of observed avoided level crossings (see
Table~\ref{tab:Expparameters}). }\label{fig:decayrate}
\end{figure}


\clearpage
\begin{table}
\caption{Parameters of observed TLFs. N is the TLF number, $f$ is
the frequency near where the avoided level crossing occurs,
$n_{g}^{-/+}$ is the negative and positive reduced gate voltage
location of each TLF and $\Delta^{-/+}$ represents the splitting
size of the crossing at $n_{g}^{-/+}$ for the specified value
$E_{J}$. $(\varepsilon_{b}-\varepsilon_{a})$ is the asymmetry in
the potential energy of the TLF in the two wells, $Q\delta x
\cos(\eta)/e d$ is the coupling between the TLF and the CPB and
$T_{ab}$ is the tunneling term between the two potential wells.}

\begin{ruledtabular}
\begin{tabular}{c|c|c|c|c|c|c|c|c|c}
N& $f$ (GHz) & $n_{g}^{-}$& $\Delta^{-}$ (MHz) & $n_{g}^{+}$&
$\Delta^{+}$ (MHz)
& $E_{J}/k_{B}$ (K) & $(\varepsilon_{b}-\varepsilon_{a})/\kb$ (K)  & $Q_{TLF}\delta{x}\cos(\eta)/ed$ & $T_{ab}/\kb$ (K)\\
\hline 1& 20  & -0.60& 30 & 0.63 & 10 & 0.3 & 0.75 & -0.032 & 0.3 \\
\hline
2&34  & -0.43 & 140 & 0.48 & 80 & 1.0 & 1.43 & -0.074 & 0.39\\
\hline
3\footnote{The splitting size was too small to resolve.}&39 & -0.34&-&0.36 &-& 1.1 & 1.86 - 1.89 & -0.083 & $< 0.17$\\
\hline
4& 43 & -0.26 & 120 & 0.25 & 220 & 1.1 & 1.61 & 0.083 & 0.65\\
\hline
5\footnote{After annealing device at $T = 300$ K for 14 days.}& 23 & -0.57 & 90 & 0.58 & 140 & 0.55 & 0.34 & 0.078 & 0.52\\
\end{tabular}
\end{ruledtabular}
\label{tab:Expparameters}
\end{table}

\clearpage
\begin{figure}[t]
\centering \includegraphics[angle=0]{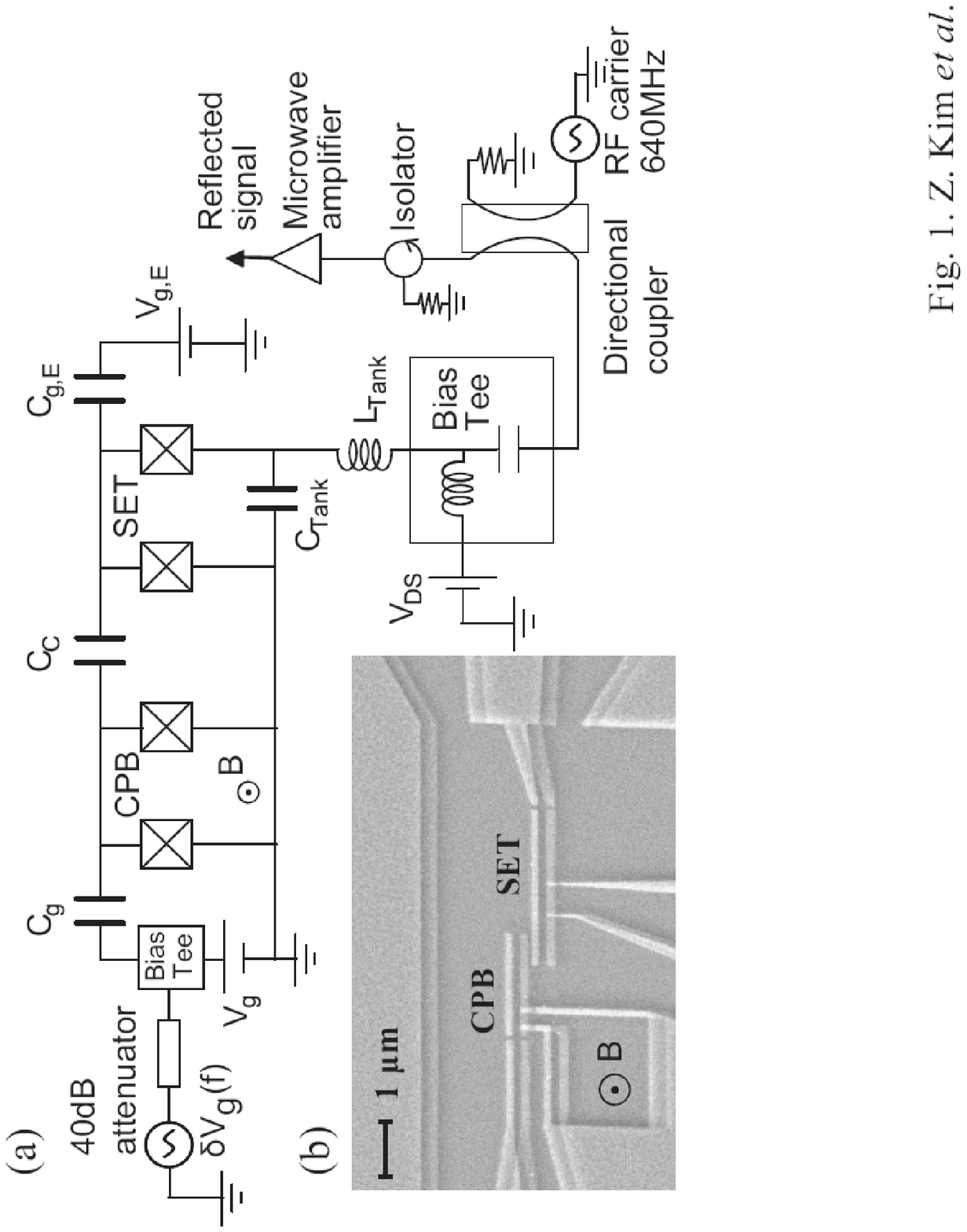}
\end{figure}

\clearpage
\begin{figure}[!t]
\centering \includegraphics[angle=0]{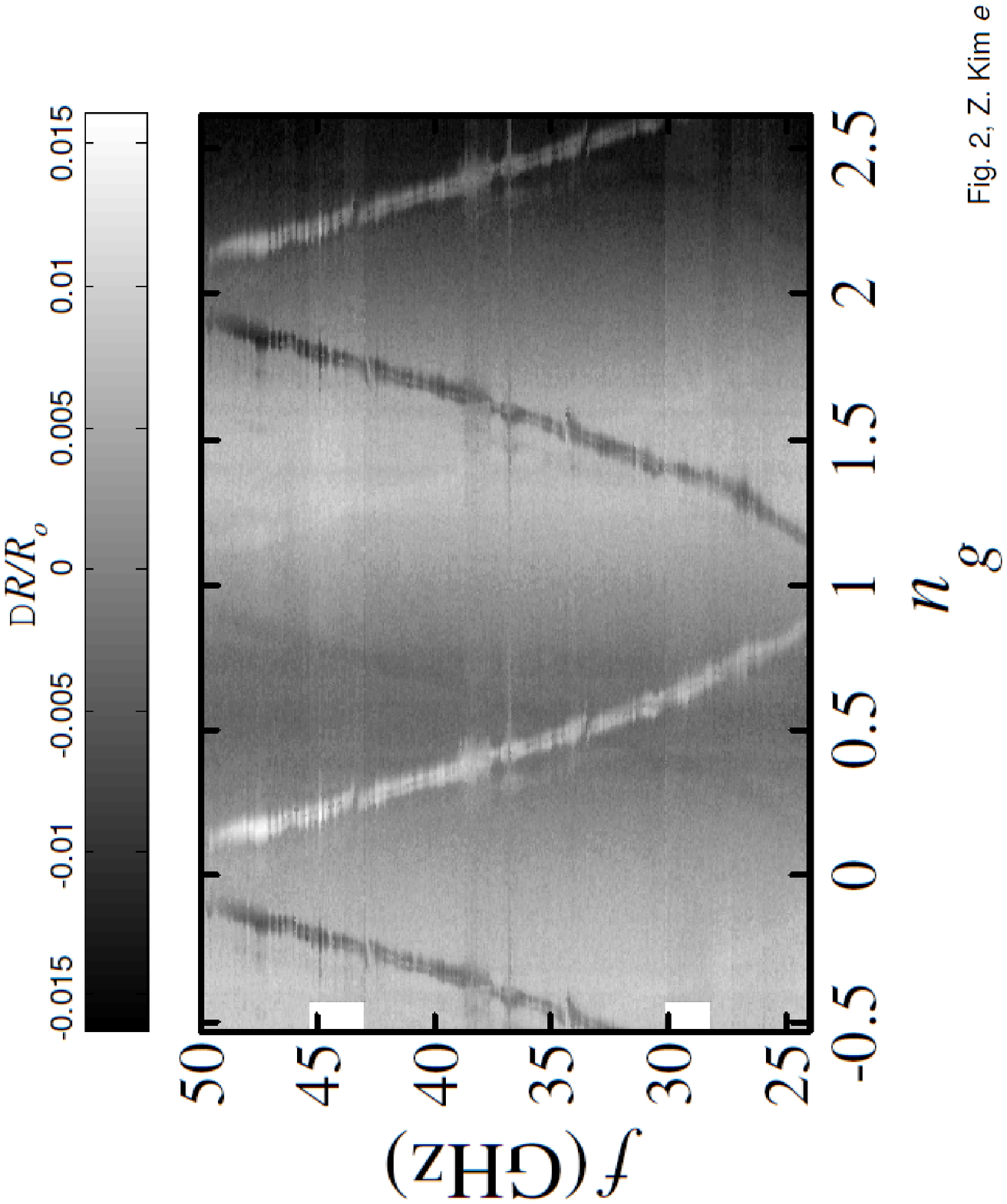}
\end{figure}

\clearpage
\begin{figure}[!t]
\centering \includegraphics[angle=0]{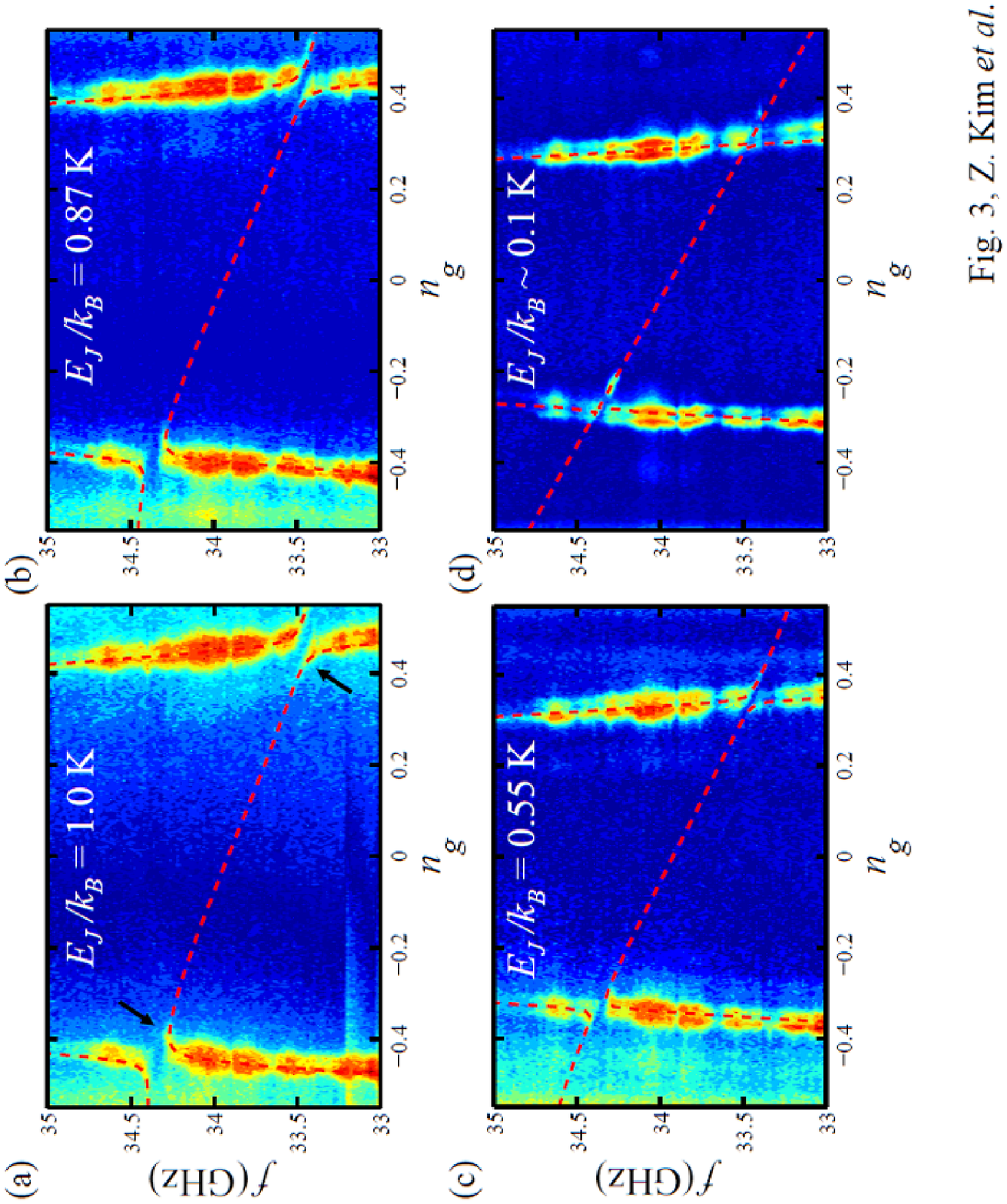}
\end{figure}

\clearpage
\begin{figure}[b]
\centering \includegraphics[angle=0]{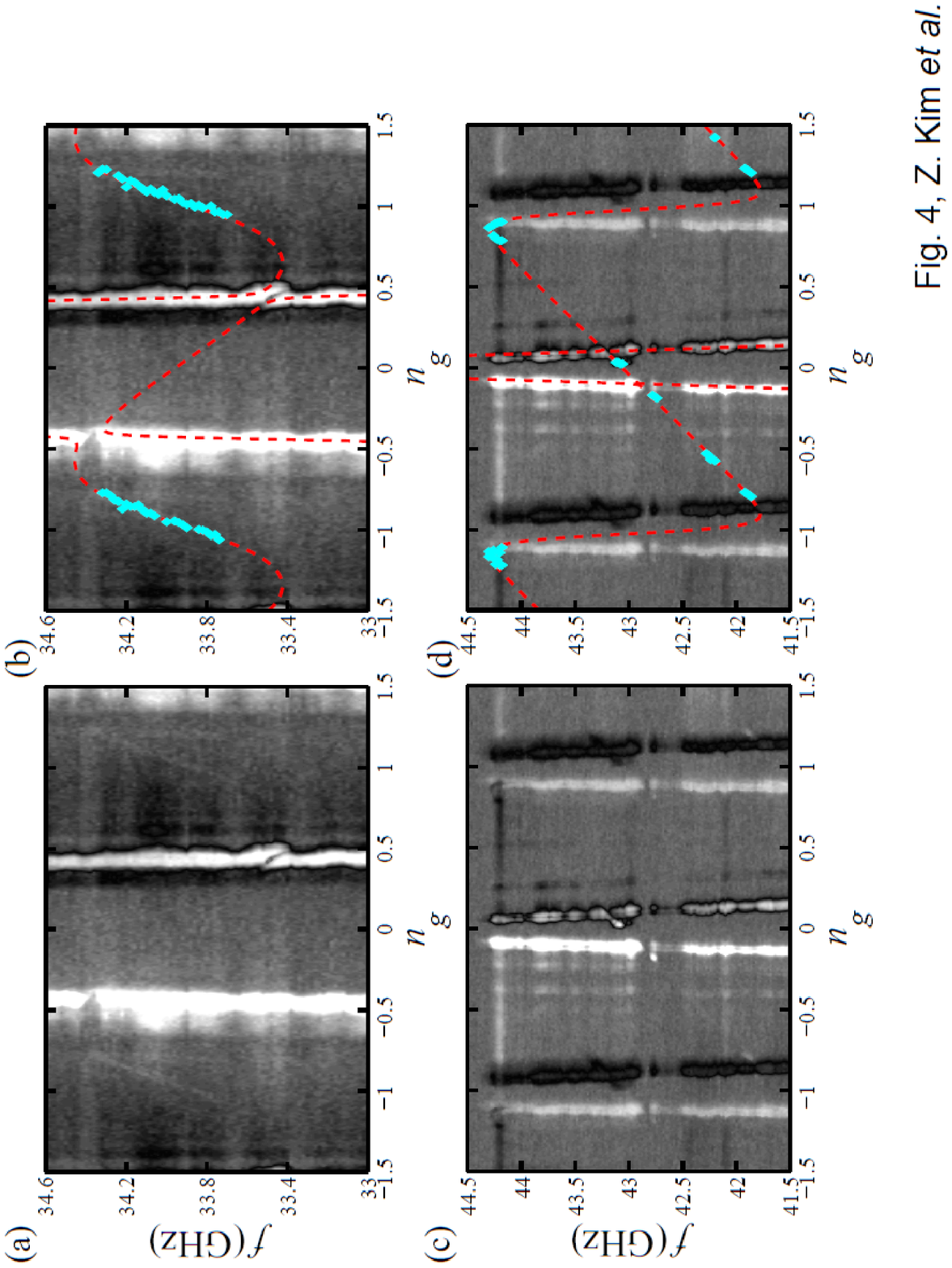}
\end{figure}

\clearpage
\begin{figure}[b]
\centering \includegraphics[angle=0]{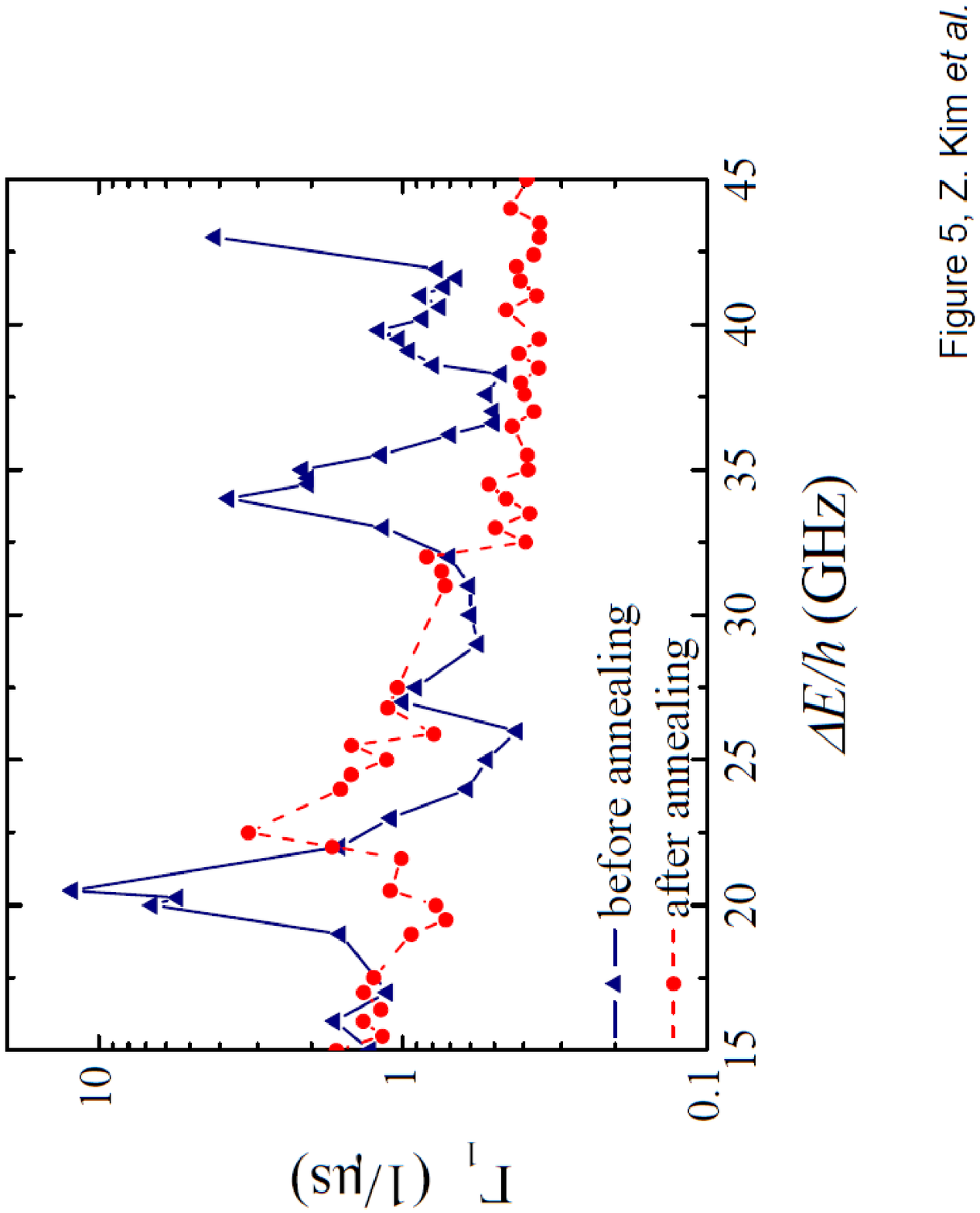}
\end{figure}

\end{document}